\documentclass[conference]{IEEEtran}
\IEEEoverridecommandlockouts
\usepackage{cite}
\usepackage{amsmath,amssymb,amsfonts}
\usepackage{algorithmic}
\usepackage{graphicx}
\usepackage{textcomp}
\usepackage{xcolor}
\usepackage{subfigure}
\usepackage{float}

\begin{document}

\title{A Phase-Coded Time-Domain Interleaved OTFS Waveform with Improved Ambiguity Function\\
}

\author{\IEEEauthorblockN{Jiajun Zhu\textsuperscript{1}, Yanqun Tang\textsuperscript{1*}, Chao Yang\textsuperscript{2}, Chi Zhang\textsuperscript{1}, Haoran Yin\textsuperscript{1}, Jiaojiao Xiong\textsuperscript{1}, Yuhua Chen\textsuperscript{1}}
\IEEEauthorblockA{\textit{\textsuperscript{1}School of Electronics and Communication Engineering, Sun Yat-sen University, Shenzhen, China} \\
\textit{\textsuperscript{2}School of Automation, Guangdong University of Technology, Guangzhou, China}\\
Email: zhujj59@mail2.sysu.edu.cn, \textsuperscript{*}tangyq8@mail.sysu.edu.cn}
}
\maketitle
\begin{abstract}

Integrated sensing and communication (ISAC) is a significant application scenario in future wireless communication networks, and sensing capability of a waveform is always evaluated by the ambiguity function.
To enhance the sensing performance of the orthogonal time frequency space (OTFS) waveform, we propose a novel time-domain interleaved cyclic-shifted P4-coded OTFS (TICP4-OTFS) with improved ambiguity function.
TICP4-OTFS can achieve superior autocorrelation features in both the time and frequency domains by exploiting the multicarrier-like form of OTFS after interleaved and the favorable autocorrelation attributes of the P4 code.
Furthermore, we present the vectorized formulation of TICP4-OTFS modulation as well as its signal structure in each domain. 
Numerical simulations show that our proposed TICP4-OTFS waveform outperforms OTFS with a narrower mainlobe as well as lower and more distant sidelobes in terms of delay and Doppler-dimensional ambiguity functions, and an instance of range estimation using pulse compression is illustrated to exhibit the proposed waveform’s greater resolution. Besides, TICP4-OTFS achieves better performance of bit error rate for communication in low signal-to-noise ratio (SNR) scenarios. 
\end{abstract}


\section{Introduction}
The development of next-generation wireless networks has highlighted the increasing importance of sensing capabilities, paving the way for the introduction of integrated sensing and communication (ISAC) \cite{ref1}. Waveform design is an important part of ISAC research since it allows for the seamless integration of radar and communication features on a single hardware device, facilitating spectrum sharing and enabling synergistic collaboration to uncover greater capabilities \cite{ref2}. Designing communication waveforms to improve their sensing performance is one of the schemes in ISAC waveform design.

Orthogonal frequency division multiplexing (OFDM) has received extensive usage as a communication waveform in commercial fourth and fifth generation wireless networks, and some research on exploiting OFDM for ISAC applications has been undertaken \cite{refISACOFDM1}. 
However, in cases requiring high-speed movement, OFDM has limitations, necessitating the investigation of alternate waveforms. 
In this regard, orthogonal time frequency space (OTFS) has emerged as a viable method capable of minimizing the negative consequences of significant Doppler shifts. The increased interest in OTFS from academia and industry emphasizes it as a promising candidate for next-generation wireless networks \cite{refOTFS2}.
Notably, the distinguishing aspect of OTFS is its parameterization via delay and Doppler indices, which aligns well with the physical interpretation of radar targets in terms of range and velocity. These intrinsic properties make OTFS a good fit for ISAC applications, as demonstrated by \cite{refOTFS1,refOTFS2,refOTFS3,refOTFS4}.

Radar sensing is a key component of ISAC, and some work has focused on radar sensing applications for OTFS based on matched filter algorithms \cite{OTFSRADARmf1, OTFSRADARpdmf2, OTFSRADARpdmf3, refOTFSaf1}. Moreover, since matched filtering is one of the most important methods in radar sensing, it is crucial to study the ambiguity function because the sensing performance in terms of e.g., the resolution performance of velocimetry and ranging is characterized by the ambiguity function \cite{yuan2023new}.
\cite{OTFSRADARmf1} proposed an efficient OTFS-based matched filter algorithm for target range and velocity estimation, which allows longer range radar and larger Doppler frequency estimation. \cite{OTFSRADARpdmf2} proposed a framework to achieve pulse Doppler radar processing using an existing OTFS communications architecture, based on which \cite{OTFSRADARpdmf3} proposed a two-dimensional correlation-based algorithm to estimate the fractional delay and Doppler parameters for radar sensing. However, there has been little research into the OTFS ambiguity function in the context of sensing performance. In a related study \cite{refOTFSaf1}, the authors proposed a random-padded OTFS modulation scheme for joint communication and radar/sensing systems. 
This OTFS variation increased radar system efficiency by drastically lowering sidelobes in the ambiguity function of the transmitted signal. It should be noted, however, that the work was designed particularly for a zero-padded OTFS signal, making it inapplicable to OTFS waveforms that do not use the zero-padded form. 

In this paper, we focus on improving the ambiguity function of the OTFS waveform on the transmitter side since the ambiguity function is determined only by the transmit waveform, from which the resolving power, ambiguity, measurement accuracy, and clutter suppression capability of the radar system under optimal signal processing conditions can be known. In another word, the ambiguity function serves as a guide for selecting radar waveforms for different purposes and characterises the applicability and limitations of a certain signal, and it is significant for ISAC systems based on specific waveform.
Specifically, we commence our investigation by examining the alignment of the OTFS frame in the time domain and analyzing the corresponding forms of the Zak transform in both analog and digital domains.
Notably, the OTFS waveform, which employs the inverse discrete Zak transform (IDZT) \cite{DD1}, \cite{refZAK2}, has a multicarrier structure similar to that of the OFDM waveform, which employs the inverse discrete Fourier transform (IDFT). 
This observation motivates us to investigate the OTFS signal's suitability for techniques used in frequency modulated waveforms.
We demonstrate a link between the ambiguity function and the signal's autocorrelation features. Based on this insight, we propose using a row-column interleaver in the digital time domain to change the alignment of the OTFS waveform, hence improving its autocorrelation features. In addition, we use a phase coding approach that is extensively used in radar waveforms. We encode delay-Doppler symbols with P4 code sequences \cite{P4} Doppler-wise and cyclically shift them by a value corresponding to the Doppler index, similar to the encoding mode in multifrequency complementary phase (MCPC) schemes \cite{MCPC1}. 
Finally, we generate the vectorized form of the proposed time-domain interleaved cyclic-shifted P4-coded OTFS (TICP4-OTFS) to assist future investigation of this waveform.

Through numerical simulations, we demonstrate that the strategy proposed in this paper produces a superior ambiguity function for radar sensing while also enhancing the bit error rate (BER) performance in low signal-to-noise ratio (SNR) scenarios for communication applications.
In addition, we illustrate an instance of range estimation using pulse compression, exhibiting the proposed waveform's greater resolution.

\section{Fundamental concepts}
In this section, we present the signal model for Zak-based OTFS and another equivalent implementation based on OFDM \cite{DD1}. Additionally, the definition of the ambiguity function and the P4 code are introduced.
\subsection{Signal model of OTFS}

In the subsequent analysis, we introduce a time-domain signal $s(t)$ with a time limit of $T_w=NT$ and a band limit of $B_w=M \Delta f$, where $\Delta f$ is set to $1/T$ to ensure orthogonality in the time-frequency domain.
By employing the Zak transform, we can express the Zak domain signal of $s(t)$ within the fundamental region of the delay-Doppler domain, denoted as $\mathcal R=\{\tau\in[0,T),\nu\in[0,\Delta f)\}$, as
\begin{equation}
Z_T[s(t)](\tau,\nu)=\sqrt T \sum_{ n=0}^{N-1}s(\tau+ nT)e^{-j2\pi nT\nu},\label{eq1}
\end{equation}
and the corresponding inverse transform of Zak is defined as
\begin{equation}
s(\tau+ nT)=\sqrt T\int_0^{\Delta f}Z_T[s(t)](\tau+nT,\nu)e^{j2\pi nT\nu}d\nu,\label{eq2}
\end{equation}
where $ n=0,1,\dots,N-1$. After sampling time $t$ and delay $\tau$ at intervals of $1/B_w=T/M$ and sampling Doppler at intervals of $1/T_w=1/(NT)$, we obtain the indices $q=t M/T$ for $q=0,1,\dots,NM-1$, $l=\tau M/T$ for $l=0,1,\dots,M-1$, and $k=\nu NT$ for $k=0,1,\dots,N-1$. Consequently, the digital time-domain signal model can be expressed as
\begin{equation}
s[q]=s[l+nM]=\frac{1}{\sqrt N}\sum_{k=0}^{N-1}Z[l,k]e^{j2\pi\frac {kn}{N}},\label{eq3}
\end{equation}
where $Z[l,k]$ represents the original information symbols located at the $l$-th delay and $k$-th Doppler of the delay-Doppler plane grid $\Gamma:\{l\frac{T}{M},k\frac{1}{NT}\}$ for $l=0,1,\dots,M-1$ and $k=0,1,\dots,N-1$. Additionally, $s[q]=s[l+nM]$ denotes the time samples corresponding to the $l$-th delay and the $n$-th time slot of the delay-time plane after column-wise parallel to serial (P/S) conversion.

Revisiting another implementation of OTFS in the digital domain, which is based on OFDM and proven to be equivalent to IDZT-based OTFS. Firstly, the delay-Doppler domain symbols $Z[l,k]$ are mapped into the time-frequency domain using inverse symplectic finite Fourier transform (ISFFT) as

\begin{equation}
X[m,n]=\frac{1}{\sqrt{MN}}\sum_{l=0}^{M-1}\sum_{k=0}^{N-1}Z[l,k]e^{j2\pi(\frac{kn}{N}-\frac{lm}{M})},\label{eq4}
\end{equation}
where $X[m,n]$ represents information symbols located at the $m$-th frequency and $n$-th time of the time-frequency plane grid $\Pi:\{nT,\frac{m}{T}\}$ with $m=0,1,\dots,M-1$, $n=0,1,\dots,N-1$. 

Then, using IDFT the time-frequency symbols $X[m,n]$ are mapped into the delay-time domain. Finally, the delay-time symbols $S[l,n]$ are rearranged in a column-wise P/S conversion to obtain the time-domain symbols $s[q]=s[l+nM]$. The process can be denoted as
\begin{equation}
\begin{aligned}
&S[l,n]=\frac{1}{\sqrt{M}}\sum_{m=0}^{M-1}X[m,n]e^{j2\pi\frac{ml}{M}},\\
&S[l,n]\xrightarrow{\rm{P/S}} s[l+nM]=s[q].\label{eq5}
\end{aligned}
\end{equation}

\subsection{Ambiguity function and P4 code}
To analyze the performance and characteristics of the waveform in greater detail, we introduce the concept of the ambiguity function. The ambiguity function serves as a metric to evaluate the sensing performance of a waveform, revealing the interference caused by a transmitted signal due to variations in delay $\tau$ and Doppler shift $\nu$ compared to a reference signal. In this paper, the ambiguity function is defined as
\begin{equation}
\chi (\tau,\nu)=\int_{-\infty}^{+\infty}s(t)s^{*}(t+\tau)e^{j2\pi \nu t}dt.\label{eq8}
\end{equation}

If we set $\nu=0$ or $\tau=0$, the delay and Doppler-dimensional versions of the ambiguity function can be obtained, respectively. The delay and Doppler-dimensional ambiguity function is defined as
\begin{equation}
\chi (\tau)=\int_{-\infty}^{+\infty}s(t)s^{*}(t+\tau)dt,\label{eqAFt}
\end{equation}

\begin{equation}
\chi (\nu)=\int_{-\infty}^{+\infty}S(f)S^{*}(f+\nu)df,\label{eqAFf}
\end{equation}
where $s(t)$ and $S(f)$ denote the signal in the time and frequency domain, respectively, and $(\cdot)^*$ is the conjugate operation.
It can be noted that the definition of the ambiguity function in \eqref{eqAFt} and \eqref{eqAFf} is the same as the autocorrelation function. Therefore, improving the autocorrelation properties of the signal in both the time and frequency domains is crucial to optimizing the sensing performance.

In terms of the P4 code, it has good autocorrelation properties in both the time and frequency domains for inheriting the advantages of the linear frequency modulation (LFM) waveform, which has the potential to be used to improve the ambiguity function of waveforms. The P4 code is generated conceptually by downconverting the LFM waveform to baseband using a local oscillator and then sampling it at the Nyquist rate, so that the successive samples are given by
\begin{equation}    
\phi_p = \pi(p-1)^2/P-\pi(p-1),\label{eq21}
\end{equation}
where $P$ denotes the number of samples and $p=1,2,\dots,P$. 

According to the definition of conventional OTFS and ambiguity function, the autocorrelation performance of OTFS in time and frequency domains is lackluster due to its dispersed time-domain sample arrangement and the shape of its envelope superimposed on a sinusoidal wave.


\begin{figure}[t]   
	\centering	
	\includegraphics[width=\linewidth]{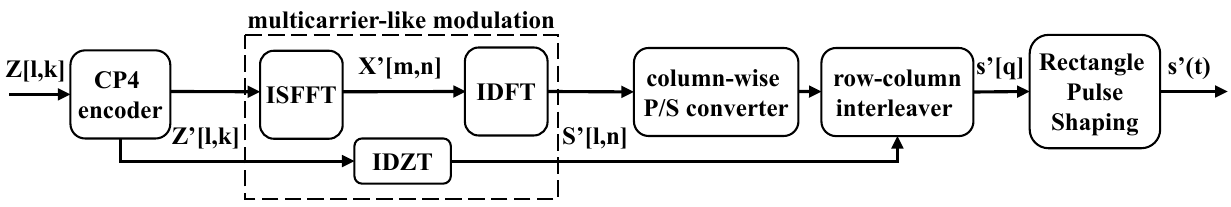}	
	\caption{Block diagram of proposed TICP4-OTFS modulation}	
	\label{block}	
\end{figure}
\section{Proposed TICP4-OTFS}
In this section, we introduce the TICP4-OTFS signal model and derive the vectorized formulation at the transmitter side. This enables us to describe the structure of the TICP4-OTFS signal in each domain, emphasizing its distinct arrangement and properties.

\subsection{Signal model of TICP4-OTFS}
By observing the form of Zak-based OTFS in the digital domain, we can identify an intrinsic form of multicarrier modulation. Equation \eqref{eq3} can be seen as an IDFT for the Doppler indices $k$, with the results parameterized by the time indices $n$, if we treat the delay indices $l$ as constant. This observation implies that the delay-Doppler plane can be viewed as a time-frequency plane with different inter-symbol and inter-carrier spacing, where the carriers are the Doppler indices, thus indicating another paradigm of multicarrier modulation. 

For all $N$ subcarriers in the proposed TICP4-OTFS signal, we use cyclic-shifted P4 phase code sequences. Each subcarrier corresponds to $M$ symbols with different delay indices in the delay-Doppler plane grid. Each P4 code sequence, which is made up of $M$ samples, is multiplied by $M$ delay-domain symbols and cyclically shifted using different Doppler indices $k$. Additionally, a row-column interleaver is employed to modify the alignment in the digital time domain. As a result, the TICP4-OTFS signal is formulated as
\begin{equation}
s'[q]=s'[lN+n]=\frac{1}{\sqrt N}\sum_{k=0}^{N-1}Z[l,k]e^{j\phi_{[l-k]_M}}e^{j2\pi\frac {kn}{N}},\label{eq22}
\end{equation}
where $(\cdot)_M$ represents the modulo $M$ operation. In addition to the definition in \eqref{eq22}, the TICP4-OTFS signal can also be acquired using an OFDM-based method. Firstly, using ISFFT, the time-frequency domain symbols can be obtained as
\begin{equation}
X'[m,n]=\frac{1}{\sqrt{MN}}\sum_{l=0}^{M-1}\sum_{k=0}^{N-1}Z[l,k]e^{j\phi_{[l-k]_M}}e^{j2\pi(\frac{kn}{N}-\frac{lm}{M})}.\label{eq23}
\end{equation}

Then, using IDFT, we can obtain the delay-time signal as
\begin{equation}
\begin{aligned}
S'[l,n]=\frac{1}{\sqrt{M}}\sum_{m=0}^{M-1}X'[m,n]e^{j2\pi\frac{ml}{M}},\label{delay-time}
\end{aligned}
\end{equation}

Finally, the time-domain signal $s'[q]=s'[lN+n]$ can be obtained by performing a column-wise P/S conversion followed by a row-column interleaving operation, as
\begin{equation}
\begin{aligned}
S'[l,n]\xrightarrow{\rm{P/S}} s'[lN+n]=s'[q].\label{time}
\end{aligned}
\end{equation}

In conclusion, we obtain the TICP4-OTFS samples in the digital domain by using the cyclic-shifted P4 encoder and a row-column interleaver. The analog domain signal of TICP4-OTFS is then obtained using rectangular pulse shaping, allowing us to investigate the TICP4-OTFS waveform and its ambiguity function. Fig. \ref{block} depicts the modulation process. 

\begin{figure}[t]   
	\centering	
	\includegraphics[width=\linewidth,scale=1.00]{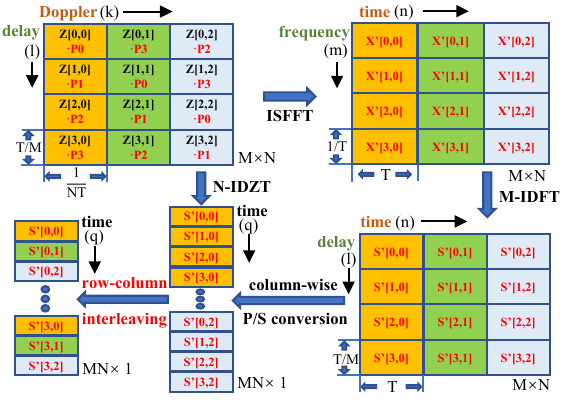}	
	\caption{Structure of proposed TICP4-OTFS}	
	\label{MDPCblock}	
\end{figure}

\begin{figure*}[htbp]   
 \begin{minipage}{0.5\textwidth}
 	\centering	
    \subfigure[Delay dimensional]{
	\includegraphics[width=0.48\linewidth]{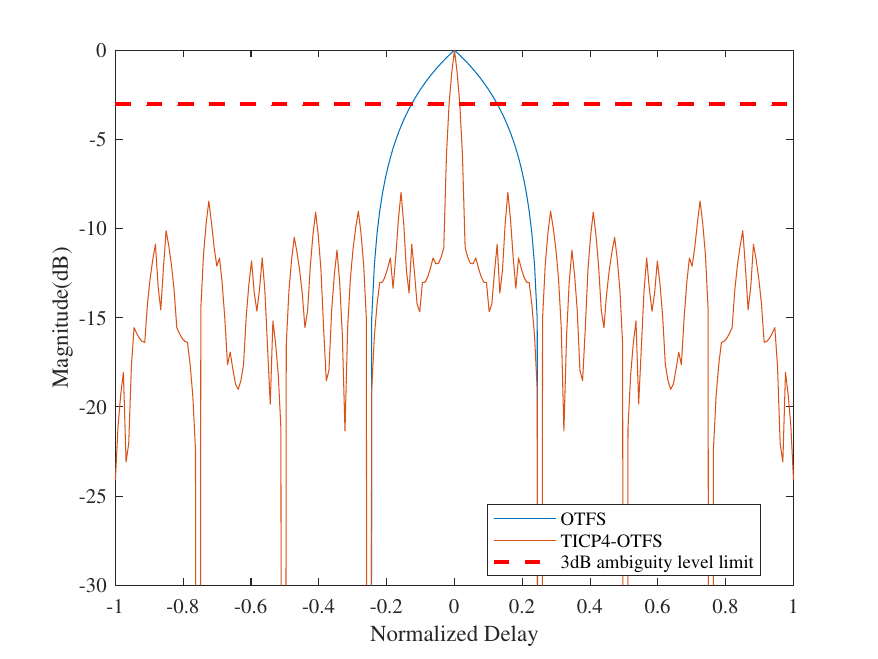}	
	\label{MDPCdelay1}	}\hspace{-3mm}
    \subfigure[Doppler dimensional]{
	\includegraphics[width=0.48\linewidth]{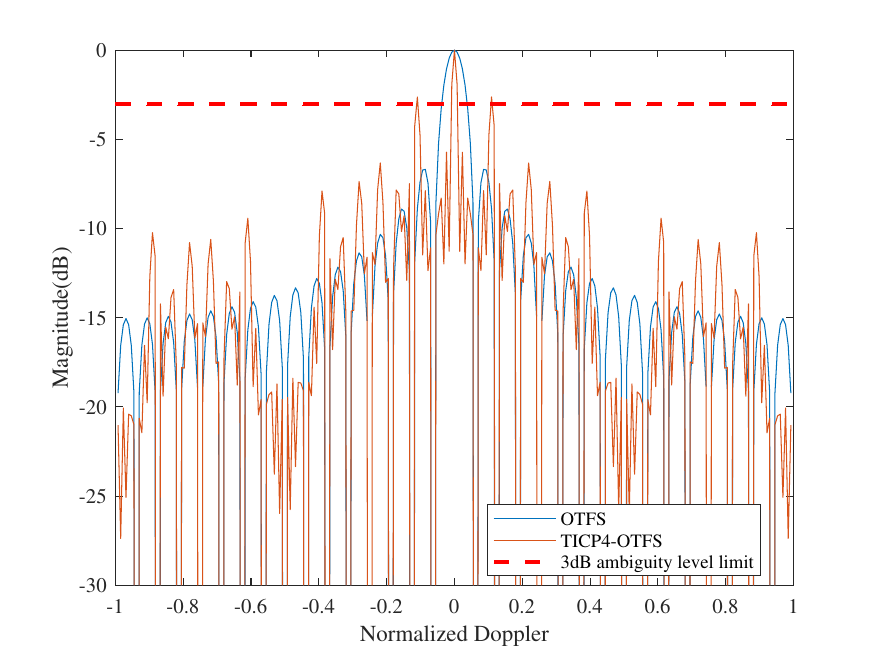}	
	\label{MDPCDoppler1}	}
 	\caption{Zero Doppler and delay-cut ambiguity function comparison, M=8, N=4}	
        \label{MDPC1}	
        \end{minipage}
  \begin{minipage}{0.5\textwidth}
        \subfigure[OTFS]{
	\includegraphics[width=0.48\linewidth]{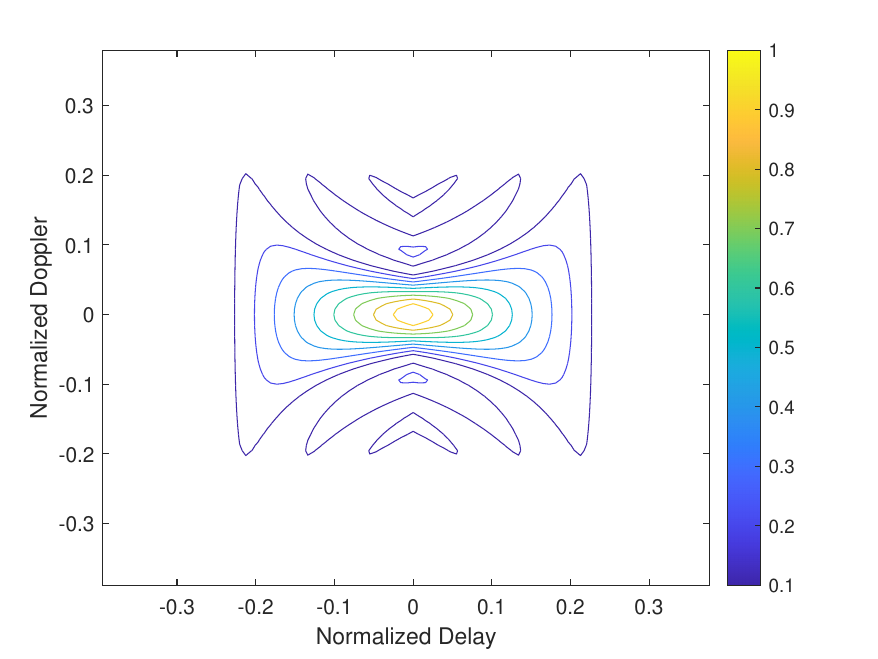}	
	\label{OTFSctr}	}\hspace{-3mm}
        \subfigure[TICP4-OTFS]{
	\includegraphics[width=0.48\linewidth]{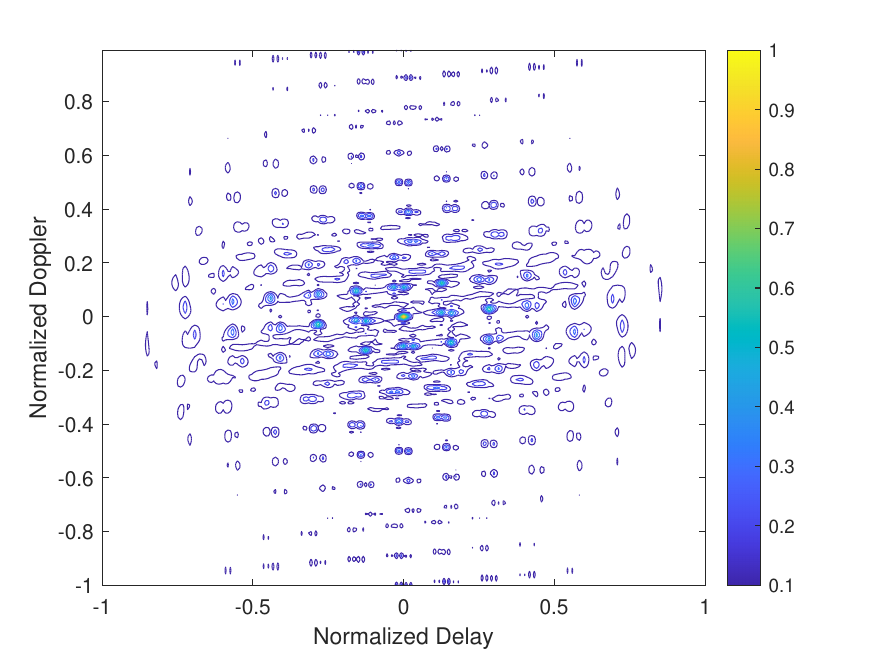}	
	\label{MDPCctr}	}
 
 	\caption{Contours of ambiguity function, M=8, N=4}	
        \label{MDPC2}	
        
        \end{minipage}
\end{figure*}

\begin{figure*}[htbp]   
 \begin{minipage}{0.5\textwidth}
 	\centering	
    \subfigure[Delay dimensional]{
	\includegraphics[width=0.48\linewidth]{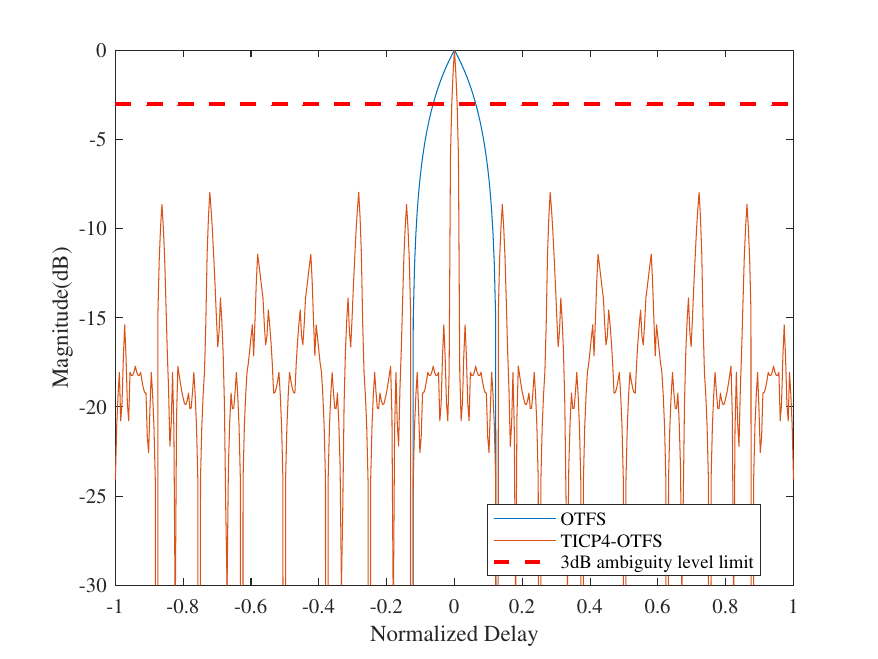}	
	\label{MDPCdelay12}	}\hspace{-3mm}
    \subfigure[Doppler dimensional]{
	\includegraphics[width=0.48\linewidth]{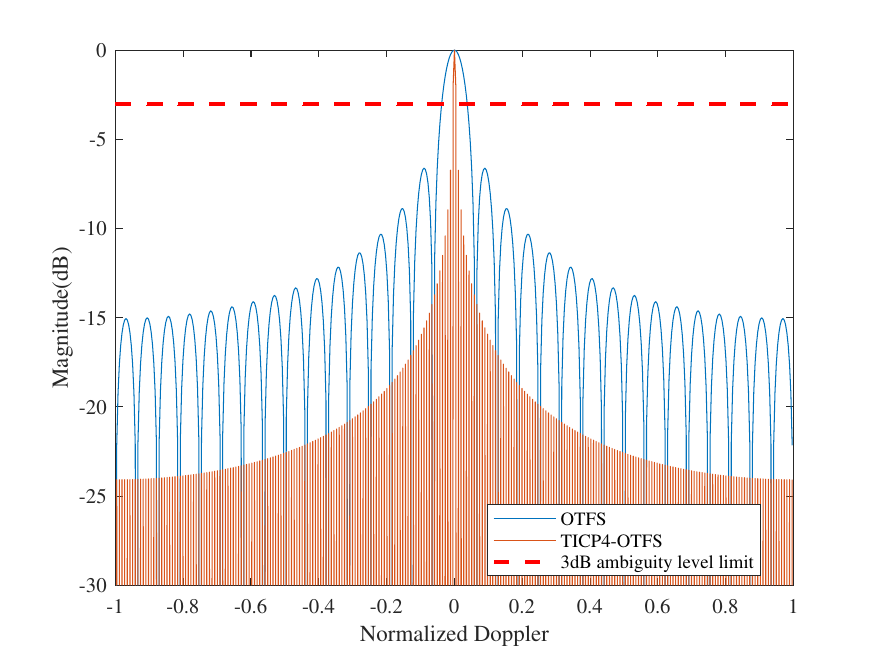}	
	\label{MDPCDoppler12}	}
 	\caption{Zero Doppler and delay-cut ambiguity function comparison, M=8, N=8}	
        \label{MDPC12}	
        \end{minipage}
  \begin{minipage}{0.5\textwidth}
        \subfigure[OTFS]{
	\includegraphics[width=0.48\linewidth]{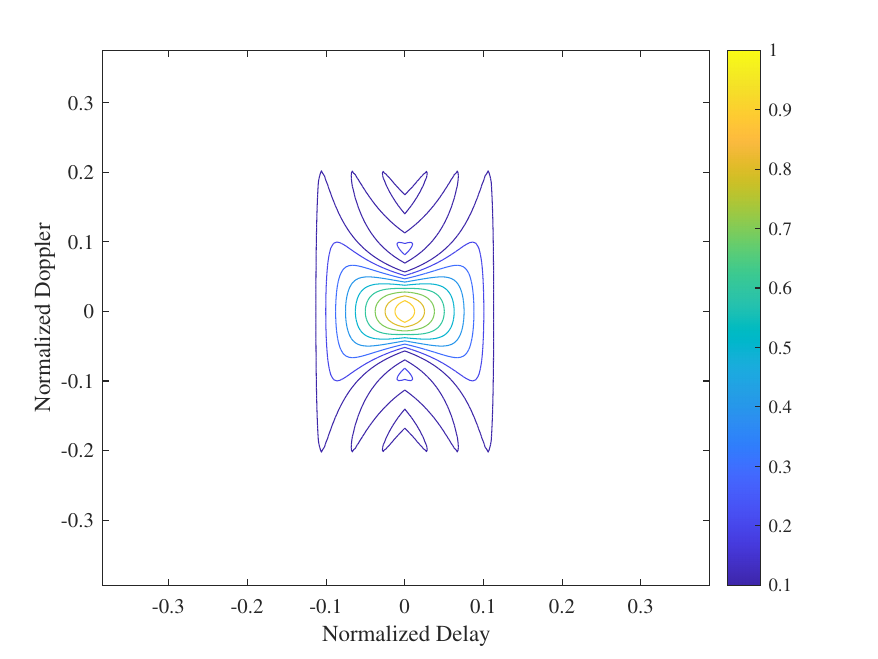}	
	\label{OTFSctr2}	}\hspace{-3mm}
        \subfigure[TICP4-OTFS]{
	\includegraphics[width=0.48\linewidth]{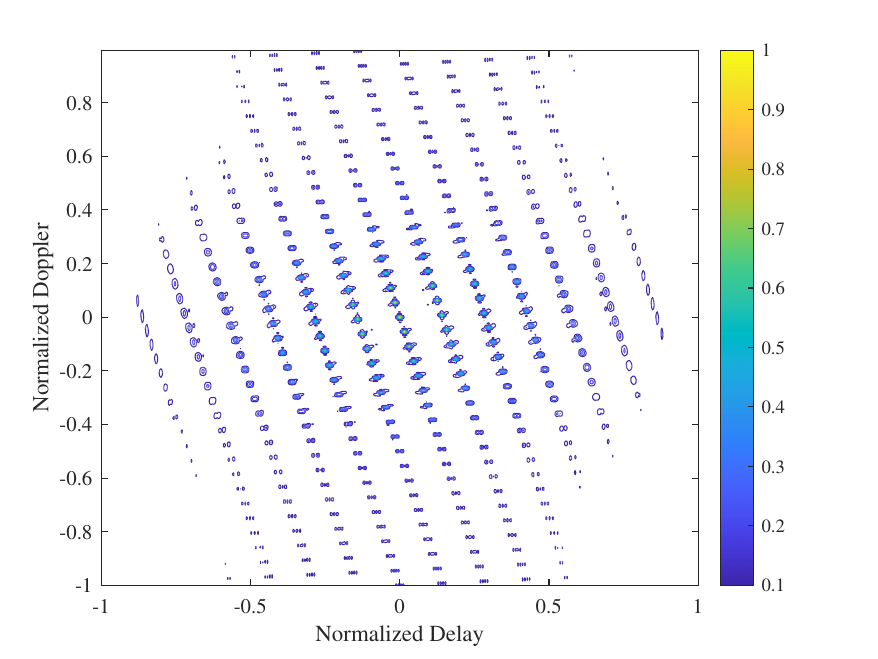}	
	\label{MDPCctr2}	}
 
 	\caption{Contours of ambiguity function, M=8, N=8}	
        \label{MDPC22}	        
        \end{minipage}
\end{figure*}
\subsection{Vectorized formulation of TICP4-OTFS modulation}
The notations used in this subsection are as follows. Uppercase boldface letters represent matrices, while lowercase boldface letters represent vectors. The Hermitian transpose is denoted by $(\cdot)^H$ and the normalized $n$-point DFT matrix is denoted by $\mathbf{F}_{\rm n}$. The identity matrix of size $k$ is denoted by the symbol $\mathbf{I}_{\rm k}$. The Kronecker and Hadamard product operations are denoted by the symbols $\otimes$ and $\odot$, respectively.

First, the transmitted signal in \eqref{delay-time} can be rewritten using a vectorized formulation as
\begin{equation}
\begin{aligned}
\mathbf S'_{\rm_{OTFS}}= \mathbf I_{\rm M}\mathbf Z'\mathbf{F_{\rm{N}}^{\rm{H}}},\label{eqOTFS}
\end{aligned}
\end{equation}
where $\mathbf Z'=\mathbf Z\odot\mathbf{\Delta_{\phi}}$ represents the input data in the delay-Doppler domain after applying the cyclic-shifted P4 code. 
$\mathbf Z$ is a $M\times N$ matrix with $Z[l,k]$ elements, and $\mathbf{\Delta_{\phi}}$ is a $M\times N$ matrix with the cyclic-shifted P4 code described as

\begin{equation}
\begin{aligned}
\mathbf{\Delta_{\phi}}&=
\begin{bmatrix}
e^{\phi_0}&e^{\phi_{M-1}}&\dots &\\ e^{\phi_1}&e^{\phi_{0}}&\dots &\\ \vdots&\vdots&\ddots&\\ e^{\phi_{M-1}}&e^{\phi_{M-2}}&\dots &
\end{bmatrix}
\\
&=
\begin{bmatrix}

P_0&P_{M-1}&\dots &\\ P_1&P_{0}&\dots &\\ \vdots&\vdots&\ddots&\\ P_{M-1}&P_{M-2}&\dots &
\end{bmatrix}.
\end{aligned}
\end{equation}

Performing column-wise vectorization on $\mathbf S_{\rm_{OTFS'}}$ yields the $MN \times 1$ vector as
\begin{equation}
\begin{aligned}
\mathbf s'_{\rm_{OTFS}}= \rm{vec}({\mathbf S'_{\rm_{OTFS}}})=(\mathbf{F_{\rm{N}}^{\rm{H}}} \otimes\mathbf I_{\rm M})\mathbf z',
\end{aligned}
\end{equation}
where $\mathbf z' =\rm{vec}(\mathbf Z' )$. To generate the proposed TICP4-OTFS signal $\mathbf s'_{\rm_{TICP4-OTFS}}$, we introduce a row-column interleaving matrix $\mathbf T\in\mathbb{C}^{NM\times NM}$ to represent the row-column interleaving operation
\begin{equation}
\mathbf T =\begin{bmatrix}
\mathbf E_{1,1}&\mathbf E_{2,1}&\dots &\mathbf E_{N,1}\\ \mathbf E_{1,2}&\mathbf E_{2,2}&\dots &\mathbf E_{N,2}\\ \vdots&\vdots&\ddots&\vdots\\ \mathbf E_{1,M}&\mathbf E_{2,M}&\dots &\mathbf E_{N,M}
\end{bmatrix},
\end{equation}
where the $N\times M$ matrix $\mathbf E_{i,j}$ is defined as
\begin{equation}
\begin{aligned}
\mathbf E_{i,j}(i',j')=\begin{cases} 1,& \text{if}\ i'=i\ \text{and}\ j'=j \\0,& \text{otherwise}  \end{cases}.
\end{aligned}
\end{equation}

Finally, we obtain the vectorized formulation of TICP4-OTFS as $\mathbf s'_{\rm_{TICP4-OTFS}}= \mathbf T\mathbf s'_{\rm_{OTFS}}$.

\subsection{Structure analysis for TICP4-OTFS signal}
Fig. \ref{MDPCblock} depicts the TICP4-OTFS transceiver, where the cyclic-shifted CP4-coded modulated symbols are mapped to the delay-Doppler domain. There are two ways for generating the time-domain TICP4-OTFS waveform for transmission over the channel. The first approach employs IDZT, which is followed by a row-column interleaver. As previously stated, IDZT allows the mapping of delay-Doppler CP4-coded symbols to the time domain, resulting in a multicarrier-like modulation. The row-column interleaver rearranges the time-domain samples while retaining all of their information.

To provide a more detailed description, we present an alternate indirect system based on OFDM. The delay-Doppler domain signal is first transferred to the time-frequency domain using an orthogonal 2D precoding approach such as the ISFFT. Following that, in each time slot, an OFDM modulator is utilized to further transform the time-frequency domain signal to the delay-time domain. Finally, the TICP4-OTFS waveform is formed by using column-wise P/S conversion followed by a row-column interleaver.

Examining the alignment of the TICP4-OTFS signal in the delay-Doppler domain and time domain reveals similarities to an OFDM signal when the delay axis is interpreted as the time axis and the Doppler axis as the frequency axis. As a result, the time domain part of TICP4-OTFS inherits the shape of the MCPC waveform, which determines the ambiguity function's features.


\section{Simulation results and discussions}
In this section, we compare the ambiguity function and BER of the conventional OTFS and TICP4-OTFS waveforms. The simulation takes into account two delay-Doppler plane grids $\Gamma$, with $M=8,8$ and $N=4,8$ for both waveforms to ensure they have the same time duration and bandwidth. 
To examine the ambiguity function, the two waveforms modulate a one-element sequence with the length of $MN$, representing a typical radar waveform without sending information. Analog time-domain waveforms are created by using rectangular pulse shaping at four times the Nyquist rate.
4-QAM modulated symbols of $MN$ are used for communication analysis. A 4-tap uniform power channel is used, with delay and Doppler taps set to $[0,1,2,3]$ and $[0,1,2,3]$, respectively. These simulations provide insight into the ambiguity function and BER performance of conventional OTFS and TICP4-OTFS waveforms.

Firstly, we compare the zero Doppler and zero delay cuts of the ambiguity function between conventional OTFS and TICP4-OTFS waveforms. Fig. \ref{MDPCdelay1} shows that the proposed TICP4-OTFS waveform has a substantially narrower main lobe in the delay dimension, around one-tenth the size of the conventional OTFS mainlobe.
Moreover, the TICP4-OTFS waveform successfully suppresses high amplitude sidelobes in the Doppler dimension, guaranteeing they are smaller in magnitude and well separated from the mainlobe, as seen in Fig. \ref{MDPCDoppler1}. 

To investigate the joint performance of the ambiguity function in the delay-Doppler domain further, we compare the two-dimensional ambiguity contours of conventional OTFS and TICP4-OTFS. 
Fig. \ref{MDPC2} clearly shows how our proposed technique transforms the ambiguity function from a column-like structure to a pegboard-like pattern. However, due to the influence of the P4 code, the TICP4-OTFS waveform may display certain coupling effects between the delay and Doppler dimensions.

Similarly, in Fig. \ref{MDPC12} and Fig. \ref{MDPC22}, we fix $N=8$ to guarantee that it corresponds to the value of $M$. TICP4-OTFS has thinner mainlobes in both the delay and Doppler dimensions of the ambiguity function than conventional OTFS. 
Furthermore, the sidelobes in the Doppler dimension are significantly suppressed compared to Fig. \ref{MDPCDoppler1}, e.g., all amplitudes except the mainlobe's are suppressed below -3dB. This enhancement could be attributed to the use of cyclic-shifted P4 codes, where the equality of the values of $M$ and $N$ prevents the Doppler-dimensional P4 codes from reoccurring. 
As a result, TICP4-OTFS achieves comparable phase coding performance in both the time and frequency domains.

\begin{figure}[t]   
	\centering	
	\includegraphics[width=\linewidth]{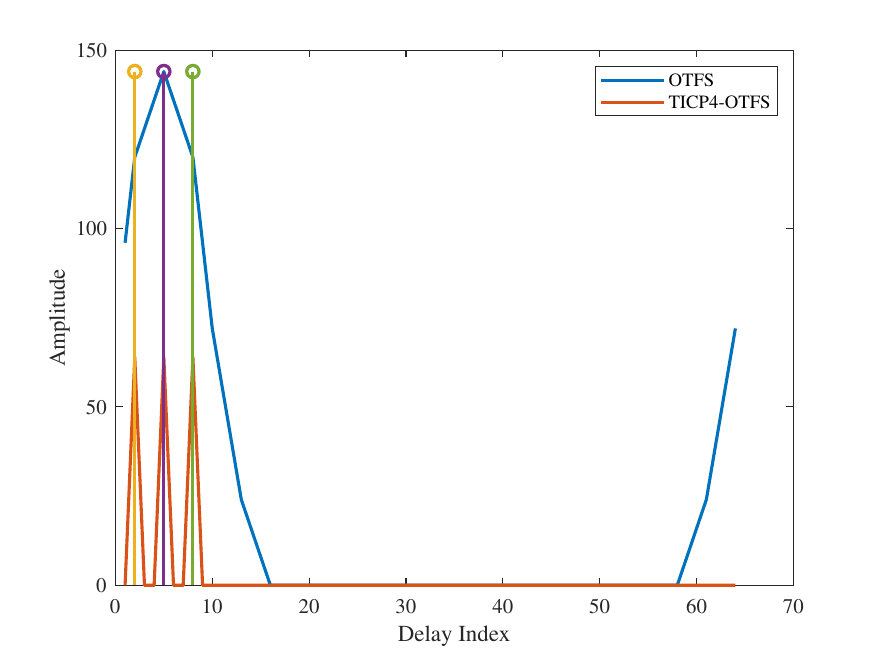}	
	\caption{Range estimation comparison, delay taps=[1,4,7]}	
	\label{RPC}	
\end{figure}

Following that, we examine the performance of two waveforms for range estimation using the pulse compression algorithm frequently used in radar by setting the delay taps to $[1,4,7]$ and the Doppler taps to $[0,0,0]$. As seen in Fig. \ref{RPC}, TICP4-OTFS exactly matches the three peaks, however OTFS mixes the three peaks together to produce a peak with a wider mainlobe due to weak resolving capacity.

\begin{figure}[t]   
	\centering	
	\includegraphics[width=\linewidth]{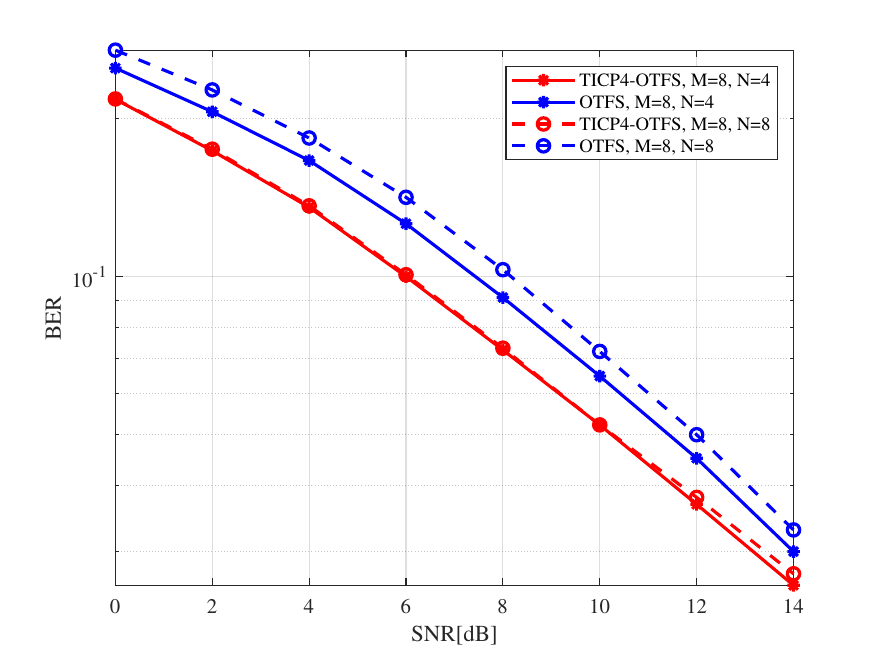}	
	\caption{BER comparison, LMMSE detection}	
	\label{BER}	
\end{figure}

Finally, we compare the BER of the two waveforms to investigate the performance of the waveforms provided in this paper in terms of communication. The simulation findings in Fig. \ref{BER} reveal that TICP4-OTFS has a lower BER than OTFS in low SNR scenarios. When combined with the ambiguity function and BER performance, we can conclude that TICP4-OTFS may be more appropriate for ISAC applications.



\section{Conclusion and future works}
In this paper, we propose a novel TICP4-OTFS to improve the ambiguity function shape. This approach leverages the multicarrier-like form of IDZT-based OTFS after interleaved and the favorable autocorrelation properties of the P4 code. 
By applying cyclic-shifted P4 code sequences in the delay-Doppler domain and incorporating a row-column interleaver in the time domain, the TICP4-OTFS waveform is generated. 
We also derive the vectorized formulation of this waveform, which serves as a valuable contribution for further exploration of efficient techniques and algorithms for TICP4-OTFS. 
Simulation results demonstrate that TICP4-OTFS exhibits a narrower mainlobe, and lower as well as more distant sidelobes while achieving better BER performance in low SNR scenarios. 
These favorable characteristics indicate the potential of TICP4-OTFS for application in ISAC scenarios, which will be investigated in our future work.


\begin{thebibliography}{10}
\providecommand{\url}[1]{#1}
\csname url@samestyle\endcsname
\providecommand{\newblock}{\relax}
\providecommand{\bibinfo}[2]{#2}
\providecommand{\BIBentrySTDinterwordspacing}{\spaceskip=0pt\relax}
\providecommand{\BIBentryALTinterwordstretchfactor}{4}
\providecommand{\BIBentryALTinterwordspacing}{\spaceskip=\fontdimen2\font plus
\BIBentryALTinterwordstretchfactor\fontdimen3\font minus
  \fontdimen4\font\relax}
\providecommand{\BIBforeignlanguage}[2]{{%
\expandafter\ifx\csname l@#1\endcsname\relax
\typeout{** WARNING: IEEEtran.bst: No hyphenation pattern has been}%
\typeout{** loaded for the language `#1'. Using the pattern for}%
\typeout{** the default language instead.}%
\else
\language=\csname l@#1\endcsname
\fi
#2}}
\providecommand{\BIBdecl}{\relax}
\BIBdecl

\bibitem{ref1}
Z.~Wei, H.~Qu, Y.~Wang, X.~Yuan, H.~Wu, Y.~Du, K.~Han, N.~Zhang, and Z.~Feng,
  ``{Integrated Sensing and Communication Signals Toward 5G-A and 6G: A
  Survey},'' \emph{IEEE Internet of Things Journal}, vol.~10, no.~13, pp.
  11\,068--11\,092, 2023.

\bibitem{ref2}
W.~Zhou, R.~Zhang, G.~Chen, and W.~Wu, ``{Integrated Sensing and Communication
  Waveform Design: A Survey},'' \emph{IEEE Open Journal of the Communications
  Society}, vol.~3, pp. 1930--1949, 2022.

\bibitem{refISACOFDM1}
C.~Sturm and W.~Wiesbeck, ``{Waveform Design and Signal Processing Aspects for
  Fusion of Wireless Communications and Radar Sensing},'' \emph{Proceedings of
  the IEEE}, vol.~99, no.~7, pp. 1236--1259, 2011.

\bibitem{refOTFS2}
Z.~Wei, W.~Yuan, S.~Li, J.~Yuan, G.~Bharatula, R.~Hadani, and L.~Hanzo,
  ``{Orthogonal Time-Frequency Space Modulation: A Promising Next-Generation
  Waveform},'' \emph{IEEE Wireless Communications}, vol.~28, no.~4, pp.
  136--144, 2021.

\bibitem{refOTFS1}
R.~Hadani, S.~Rakib, M.~Tsatsanis, A.~Monk, A.~J. Goldsmith, A.~F. Molisch, and
  R.~Calderbank, ``{Orthogonal Time Frequency Space Modulation},'' \emph{2017 IEEE Wireless Communications and Networking Conference (WCNC)}, pp. 1--6, 2017.

\bibitem{refOTFS3}
Z.~Wei, S.~Li, W.~Yuan, R.~Schober, and G.~Caire, ``{Orthogonal Time Frequency
  Space Modulation—Part I: Fundamentals and Challenges Ahead},'' \emph{IEEE
  Communications Letters}, vol.~27, no.~1, pp. 4--8, 2023.
  
 \newpage
 
\bibitem{refOTFS4}
W.~Yuan, Z.~Wei, S.~Li, R.~Schober, and G.~Caire, ``{Orthogonal Time Frequency
  Space Modulation—Part III: ISAC and Potential Applications},'' \emph{IEEE
  Communications Letters}, vol.~27, no.~1, pp. 14--18, 2023.

\bibitem{OTFSRADARmf1}
P.~Raviteja, K.~T. Phan, Y.~Hong, and E.~Viterbo, ``{Orthogonal Time Frequency
  Space (OTFS) Modulation Based Radar System},'' \emph{2019 IEEE Radar
  Conference (RadarConf)}, pp. 1--6, 2019.

\bibitem{OTFSRADARpdmf2}
K.~{Zhang}, W.~{Yuan}, S.~{Li}, F.~{Liu}, F.~{Gao}, P.~{Fan}, and Y.~{Cai},
  ``{Radar Sensing via OTFS Signaling: A Delay Doppler Signal Processing
  Perspective},'' \emph{arXiv e-prints}, p. arXiv:2301.09909, 2023.

\bibitem{OTFSRADARpdmf3}
A.~S. Bondre and C.~D. Richmond, ``{Dual-Use of OTFS Architecture for Pulse
  Doppler Radar Processing},'' \emph{2022 IEEE Radar Conference
  (RadarConf22)}, pp. 1--6, 2022.

\bibitem{refOTFSaf1}
P.~Karpovich and T.~P. Zielinski, ``{Random-Padded OTFS Modulation for Joint
  Communication and Radar/Sensing Systems},'' \emph{2022 23rd International
  Radar Symposium (IRS)}, pp. 104--109, 2022.

\bibitem{yuan2023new}
W.~Yuan, S.~Li, Z.~Wei, Y.~Cui, J.~Jiang, H.~Zhang, and P.~Fan, ``{New Delay
  Doppler Communication Paradigm in 6G Era: A Survey of Orthogonal Time
  Frequency Space (OTFS)},'' \emph{China Communications}, vol.~20, no.~6, pp.
  1--25, 2023.

\bibitem{DD1}
A.~F. Molisch, ``{Delay-Doppler Communications: Principles and Applications},''
  \emph{IEEE Communications Magazine}, vol.~61, no.~3, pp. 10--10, 2023.

\bibitem{refZAK2}
S.~K. Mohammed, R.~Hadani, A.~Chockalingam, and R.~Calderbank, ``{OTFS—A
  Mathematical Foundation for Communication and Radar Sensing in the
  Delay-Doppler Domain},'' \emph{IEEE BITS the Information Theory Magazine},
  vol.~2, no.~2, pp. 36--55, 2022.

\bibitem{P4} B. L. Lewis and F. F. Kretschmer, "Linear Frequency Modulation Derived Polyphase Pulse Compression Codes,"  \emph{IEEE Transactions on Aerospace and Electronic Systems}, vol. AES-18, no. 5, pp. 637-641, 1982.

\bibitem{MCPC1}
N.~Levanon, ``{Multifrequency Complementary Phase-Coded Radar Signal},''
  \emph{IEE Proceedings-Radar, Sonar and Navigation}, vol. 147, no.~6, pp.
  276--284, 2000.

\end{thebibliography}


\end{document}